\documentclass [10pt]{article}
\usepackage[T1]{fontenc}
\usepackage[final]{epsfig}
\usepackage{amssymb}
\usepackage{multicol}
\usepackage{graphics}
\usepackage{color}
\usepackage{ntimes}

\newcommand{\BE}{\begin{equation}}
\newcommand{\EE}{\end{equation}}

\begin{document}

\title{Traffic and the visual perception of space}
\author{\textbf{Petr \v Seba${}^{1,2,3}$}\\
\small{${}^1$ University of Hradec Kr\'alov\'e, Hradec Kr\'alov\'e
- Czech Republic}\\
\small{${}^2$ Institute of Physics, Academy of Sciences of the
Czech Republic, Prague - Czech Republic}\\
\small{${}^3$ Doppler Institute for Mathematical
Physics and Applied Mathematics,}\\ \small{Faculty of Nuclear
Sciences and Physical Engineering,
Czech Technical University, Prague - Czech Republic}}

\normalsize

\maketitle

\begin{abstract}

During the attempt to line up into a dense traffic people have
necessarily to share a limited space under turbulent conditions.
From the statistical point view it generally leads to a probability
distribution  of the distances between the traffic objects (cars or
pedestrians). But the problem is not restricted on humans. It comes
up again when we try to describe the statistics of distances between
perching birds or moving sheep herd. Our aim is to demonstrate that
the spacing distribution is generic and independent on the nature of
the object considered. We show that this fact is based on the
unconscious perception of space that people share with the animals.
We give a simple mathematical model of this phenomenon and prove its
validity on the real data that include the clearance distribution
between: parked cars, perching birds, pedestrians, cars moving in a
dense traffic and the distances inside a sheep herd.
\end{abstract}

\section{Introduction}
Everyone knows that to park a car in the city center is problematic.
The amount of the available places is limited and it has to be
shared between too many interested parties.  Birds face the same
problem when a flock tries to perch on an electric line. Similar
reasons lead also to the troubles in the highway traffic, to the
pedestrian queues and so on. For instance the unpopular transport
jams are a consequence of the car-car interaction in a regime of a
high car density.

The interaction is basically evoked by the brain activity and
mediated through the muscles (pedestrians) or accelerator/brake
pedals (cars). In both cases  it is the brain that is responsible
for the interaction. So it should be not surprising that the
phenomena observed for cars and pedestrians are similar. Even more:
the spatial perception is evolutionary very old and people share it
with animals. So we should be not surprised to find the same results
also when dealing with animal herds instead of human traffic.
Anyhow: a deeper understanding of the related processes is of
interest and will be discussed here.

The attempt to describe the intrinsic and basically  unconscious
mechanisms used by the  human brain  on the basis of their everyday
activity is not new. It goes back at least to the celebrated work of
G.K. Zipf \cite{zipf} describing the universal features of
languages. The concept was worked out in detail by the sociologist
H. A. Simon, within a set of assumptions which became known as the
Simon's model \cite{sim}, see also \cite{zan} for the recent results
in musicology.

The subjects of our study  (parking, car transport, pedestrian
dynamics, herd and flock dynamics) were treated separately in the
past. The random car parking model was  introduced by Renyi
\cite{Renyi} (see \cite{evans}, \cite{ca} for review) and compared
with the data collected on the street in \cite{rawal}, \cite{seba}.
Other models describe the car transport on highways - see for
instance \cite{kerner_synchr_2} and \cite{chowdhury2000} for review.
Another approach is used for the pedestrian dynamics utilizing
social forces - see \cite{schr}. For the dynamics of a starling
flock see \cite{starling} and \cite{starling2}

Our aim is to present a simple theory based on the visual perception
that is capable to describe all the observed facts regardless
whether they originate from cars, pedestrians or  animals. The
perception mechanism is very old and therefore shared by many
species ranging from insects to mammals. For humans it is processed
automatically and without the conscious control. We will use it to
understand the statistical properties of the distances between the
neighboring competitors (cars, pedestrians, birds and sheeps) in a
situation when the available space is limited. .

The paper is organized as follows: In the Section 2 we describe the
mathematical model and derive a one parameter family of possible
distance distributions. Section 3 contains the psychophysical
background of the distance control based on the unconsciously
evaluated time to collision with the neighbor.  The Section 4
contains the comparison of the theory with the collected data.

\section{Dividing the available space}
To illustrate the approach we focus on the spacing distribution
(bumper to bumper distances) between cars parked in parallel. The
generalization of the method to other situations will be discussed
at the end of this section.

We will assume that the street segment used for parking starts and
ends with some clear and non-transportable part unsuitable for
parking. It can be a driveway or a turning to the side street.
Otherwise the parking segment is free of any kind of  obstructions.
We will assume that it has a length $L$ and is free of any kind of
marked parking lots or park meters. So the drivers are free to park
the car anywhere in the segment provided they find an empty space to
do it. We suppose also for simplicity that all cars have the same
length $l_0$. Since many cars are cruising for parking there are not
free parking lots and a car can park only when another parked car
leaves. To simplify the further formulation of the problem and to
avoid troubling with the boundary effects we assume that the street
segment under consideration forms a circle. The car spacing
distribution is obtained as a steady solution of the repeated car
parking and car leaving process.

To park a car of the length $l_0$ one needs (due to the parking
maneuver) a lot of a length larger then $\approx 1.2 l_0$. So in a
segment of the length $L$ the number of the parked cars equals to
$N\approx[L/(1.2 l_0)]$. Denoting by $D_k$ the spacing between the
car $k$ and $k+1$ we get $ \sum_{k=1}^{N}D_k = L - N l_0$ and after
a simple rescaling finally

\begin{equation}\label{total}
 \sum_{k=1}^{N}D_k = 1.
\end{equation}

Since all lots are occupied  the number of the parked cars is
supposed to be fixed. The repeated car leaving and car parking
reshuffles however the distances $D_k$. We will treat them as
independent random variables constrained by the simplex
(\ref{total}). The distance reshuffling goes as follows: In the
first step one randomly chosen car leaves the street and the two
adjoining lots merge into a single one. In the second step a new car
parks into this empty space and splits it again into two smaller
lots. The related equations are simple. If a car leaves the
neighboring spacings - say the spacings $D_n,D_{n+1}$ - merge into a
single lot $D$:
\begin{equation}\label{cdistances}
    D=D_n+D_{n+1}+l_0.
\end{equation}
When a new car parks to $D$ it splits it into
 $\tilde D_n, \tilde D_{n+1}$:
\begin{eqnarray}
 \nonumber
  \tilde D_n &=& a(D-l_0) \\
  \tilde D_{n+1} &=& (1-a)(D-l_0).
  \label{cnewdistances}
\end{eqnarray}
where $a\in (0,1)$ is a random variable with a probability density
$q(a)$. The distribution $q(a)$ describes the parking preference of
the driver. We assume that all drivers share the same $q(a)$. (The
meaning of the variable $a$ is straightforward. For $a=0$ the car
parks immediately in front of the car delimiting the parking lot
from the left without leaving any empty space. For $a=1/2$ it parks
exactly to the center of the lot $D$ and for $a=1$ it stops exactly
behind the car on the right.) Combining (\ref{cdistances}) and
(\ref{cnewdistances}) gives finally the distance reshuffling
\begin{eqnarray}
 \nonumber
  \tilde D_n &=& a(D_n+D_{n+1}) \\
  \tilde D_{n+1} &=& (1-a)(D_n+D_{n+1}).
  \label{cnewdistances2}
\end{eqnarray}
and the car length $l_0$ drops out. The simplex (\ref{total}) is of
course invariant under this transformation.

For various choices of $n$ the mappings (\ref{cnewdistances2}) are
regarded as statistically independent. Since all cars are equal the
joint distance probability density $P(D_1,...,D_{N})$ has to be
exchangeable ( i.e. invariant under the permutation of the
variables) and invariant with respect to (\ref{cnewdistances2}). Its
marginals $p_k(D_k)$  (the probability density of a particular
spacing $D_k$) are identical:
\begin{equation}\label{distrib}
    p_k(D_k)=p(D_k)=\int_{D_1+..+D_N=1} P(D_1,...,D_{N})
    dD_1..dD_{k-1}dD_{k+1}..dD_{N}.
\end{equation}

A standard approach to deal with  the simplex (\ref{total}) is to
take $D_k$  as independent random variables normalized by a sum:
\begin{equation}\label{norma}
    D_k=\frac{d_k}{\sum_{n=1}^N d_n}.
\end{equation}
where $d_k$ are statistically independent and identically
distributed. Moreover: it is obvious that the distribution of
$\{D_1,..,D_N\}$ is invariant under the transform
(\ref{cnewdistances2}) merely when the distribution of
$\{d_1,..,d_N\}$ is invariant. The relation (\ref{cnewdistances2})
reads for the variables $d_n$:
\begin{eqnarray}
 \nonumber
  \tilde d_n &=& a(d_n+d_{n+1}) \\
  \tilde d_{n+1} &=& (1-a)(d_n+d_{n+1})
  \label{cnewdistances3}
\end{eqnarray}
 ($a,d_n,d_{n+1}$ are now statistically independent and
$d_n,d_{n+1}$ are identically distributed).

The parking maneuver is regarded as known and described by the
distribution $q(a)$. For simplicity we assume a symmetric maneuver,
i.e. $q(a)=q(1-a)$.  This means that the drivers are
 not biased to park more closely to a car adjacent from the behind
or from the front. With given $q(a)$ we look for the distribution of
$d_n$ that is invariant under the transform (\ref{cnewdistances3}).
In other words the effort is to solve the equation
\begin{equation}\label{perp}
     d \triangleq a(d+d')
\end{equation}
where $d'$ is an independent copy of the variable $d$ and the symbol
$\triangleq$ means that the left and right hand sides of
(\ref{perp}) have identical statistical properties.

Distributional equations of this type are mathematically well
studied - see for instance \cite{dev} - although not much is known
about their exact solutions. In particular it is known that for a
given distribution $q(a)$  the equation (\ref{perp}) has an unique
solution that can be obtained numerically by iterations. We are
however interested in an explicit result. We restrict therefore the
possible densities $q(a)$ to a two parametric class of the standard
$\beta$ distributions. Then the solution of (\ref{perp}) results
from the following statement \cite{duf1}:
\\
\\
\textbf{Statement:} Let $d_1,d_2$ and $a$ be independent random
variables with the distributions: $d_1\sim \Gamma(a_1,1)$, $d_2\sim
\Gamma(a_2,1)$ and $a\sim \beta(a_1,a_2)$. Then  $a(d_1+d_2)\sim
\Gamma(a_1,1)$.
\\
\\
The symbol $\sim$ means that the related random variable has the
specified probability density.  $\Gamma(g,1), \beta(g_1,g_2)$ denote
the standard gamma and beta distributions respectively.

For a symmetric parking maneuver $g_1=g_2=g$. The variables
$d_1,d_2$ are then equally distributed and the solution of
(\ref{perp}) reads $d\sim\Gamma(g,1)$. The relation (\ref{norma})
returns the spacings $D_k$. We find that the joint probability
density $P(D_1,...,D_{N})$ is nothing but a one parameter family of
the multivariate Dirichlet distributions on the simplex
(\ref{total}) \cite{wilks}:
\begin{equation}
\label{dirichlet}
P(D_1,...,D_{N})=\frac{\Gamma(Ng)}{\Gamma(g)^N}D_1^{g-1}
D_2^{g-1}...D_{N}^{g-1}
\end{equation}
Its marginal (\ref{distrib}) is simply $D \sim \beta(g,(N-1)g)$.
Normalizing  the mean of $D$ to $1$ we are finally left with
\begin{equation}
\label{distrib2}
    p(D)=\frac{1}{N}\beta\left(g,(N-1)g,\frac{D}{N}\right).
\end{equation}

A similar reasoning applies also for the moving cars. Assume a
sequence of cars in a high density traffic. All the cars move with
similar velocity and with the mutual distances $D_k$. In course of
the traffic flow the driver of the car $k$ tries to optimize his
position.  He can overtake the car $k+1$ or update the distances to
the neighboring cars $k-1$ and $k+1$.  In doing so he uses the same
mechanism as for parking. The only difference is that in the traffic
flow we assume the existence of certain safety margin. So the update
mapping reads now
\begin{eqnarray}
 \nonumber
  \tilde D_n &=& d+ a(D_n+D_{n+1}-2d) \\
  \tilde D_{n+1} &=& d+ (1-a)(D_n+D_{n+1}-2d).
  \label{cnewdistances4}
\end{eqnarray}

where $d$ is the safety margin representing the minimal distance.
For maneuvers related with the approach to a standing object (like
the parking maneuver) we set $d=0$. In the course of a car-following
in a dense traffic the value of $d$ reflects the reaction time of
the driver and his velocity. Changes in the driving situation like
night versus day conditions affect this component. We will however
neglect this fact and regard $d$ as a constant. For experimental
results describing the value of $d$ under various conditions  see
for instance \cite{li}.

We assume that outside the safety margin the driving strategy is
identical with that of the parking. So in a steady traffic flow the
distribution of the distances $D_k-d$ has to conforms with the
distribution obtained for the car parking.

The cars are of course not somehow extraordinary. The same reasoning
applies for pedestrians, animals in a herd and so on. In all cases
the distribution $q(a)$ is crucial. We will argue that $q(a)$ is
related to the inborn visual perception of the distance and
independent on the "hardware" actually used to realize the motion.

\section{Distance perception}
The ranging maneuver is described by the probability density $q(a)$
 and defined  by the relation
(\ref{cnewdistances2}). To ensure a solvability we restricted the
possible maneuvers to $q(a) = \beta(g,g,a)$, with $g$ being  a free
parameter. We will now demonstrate that the natural choice is $g=3$.

The point is that for small $a$ the behavior of $q(a)$  reflects the
capability of the driver/pedestrian/animal to estimate small
distances. The collision avoidance during the ranging is guided
visually. We assume that the same visual ability is shared by all
participants. If this applies the behavior of $q(a)$ for small $a$
has to be generic, i.e. independent on the particular situation. It
is fixed merely by the perception of the distance.

A distance perception is a complex task and there are several cues
for it. Some of them are monocular (linear perspective, monocular
movement parallax etc.), others oculomotor (accommodation
convergence) and finally binocular (i.e. based on the stereopsis).
In human all of them work simultaneously and are reliable under
different conditions - see \cite{jacobs} for more details. For the
ranging however the crucial information is not the distance itself
but the estimated time to collide with the neighbor which has to be
evaluated using the knowledge of the mutual distance and velocity.

It has been argued in a seminal paper by Lee \cite{lee} that the
estimated time to collision is psychophysically derived using a
quantity defined as the inverse of the relative rate of the
expansion of a retinal image of the moving object (this rate is
traditionally denoted as $\tau$). Behavioral experiments have
indicated that $\tau$ is indeed controlling actions like contacting
surfaces by flies, birds and mammals (including humans): see
\cite{weel},\cite{hop},\cite{shra}. Moreover the studies have
provided abundant evidence that $\tau$ is processed by specialized
neural mechanisms in the the retina itself and in the brain
\cite{far}. The hypothesis is that $\tau$ is the informative
variable for the collision free motion - see \cite{fajen} for
review.

Let $\theta$  be the instantaneous angular size of the observed
object (for instance the front of the car we are backing to during
the parking maneuver). Then the estimated time to contact is given
by

\begin{equation}\label{tau}
     \tau=\frac{\theta}{d\theta/dt}
\end{equation}
Since $\theta (t) =2 \arctan(L_0/2D(t))$ with $L_0$ being the width
of the approached object and $D(t)$ its instantaneous distance, we
get
\begin{equation}\label{tau2}
    \tau(t)=-\frac{L_0^2+4D(t)^2}{2L_0 (dD(t)/dt)}
    \arctan\left(\frac{L_0}{2D(t)}\right).
\end{equation}

For $D>>L_0$ and a constant approach speed $v=-dD/dt$ the quantity
$\tau$ simply equals to the physical arrival time: $\tau=D/v.$  For
small distances, however, $\tau\approx D^2/(v L_0 )$ and the
estimated time to contact decreases \it quadratically \rm with the
distance. (Note that $\tau$ gives the arrival time without
explicitly knowing the mutual velocity, the size of the object and
its distance.)

We assume that the probability to exploit small distances is \it
proportional to the estimated time to contact \rm. This means in
particular that if $\tau$ evaluated in the course of an approach is
small (i.e. a collision is impending) the maneuver is stopped. Based
on this principle we get for small distances $p(D)\approx D^2$ and
from \ref{cnewdistances} finally $q(a)\approx a^2$ for small $a$.
Since $q(a) = \beta(g,g,a)$ this sets the parameter $g$ to $g=3$.
The normalized clearance distribution (\ref{distrib2}) reads simply
\begin{equation}\label{clear}
p(D)=\frac{1}{N}\beta\left(3,3(N-1),\frac{D}{N}\right)=
\left(\frac{1}{N}\right)^{3(N-1)}\frac{\Gamma(3N)}{2\Gamma(3(N-1))}D^2(N-D)^{3N-4}
\end{equation}

The described  mechanism works so to say in the background, i.e.
without being conscious.  Moreover: $\tau$ is evaluated equally by
humans and by animals. We will show in the next section that this
fact leads to an universality in their behavior.

\section{The measured data}
The estimation of the distance $D$ through $\tau$ leads simply to
$p(D)\approx D^2$ for small $D$.  So let us first check the validity
of this relation. There exist a simple observation that enables us
to do it: the car stopping on a crossing equipped with traffic
lights. If the light is red the cars stop and form a queue. We
assume that the drivers stop independently and in a distance to the
preceding car that is evaluated by $\tau$. So the clearance
statistics should give an evidence of the validity of the $\tau$
hypothesis. There was also a direct experiment measuring the
clearance statistics in laboratory conditions - see
\cite{gadgil},\cite{GREEN}.

\begin{figure}
\begin{center}
  \includegraphics[height=9cm,width=15cm]{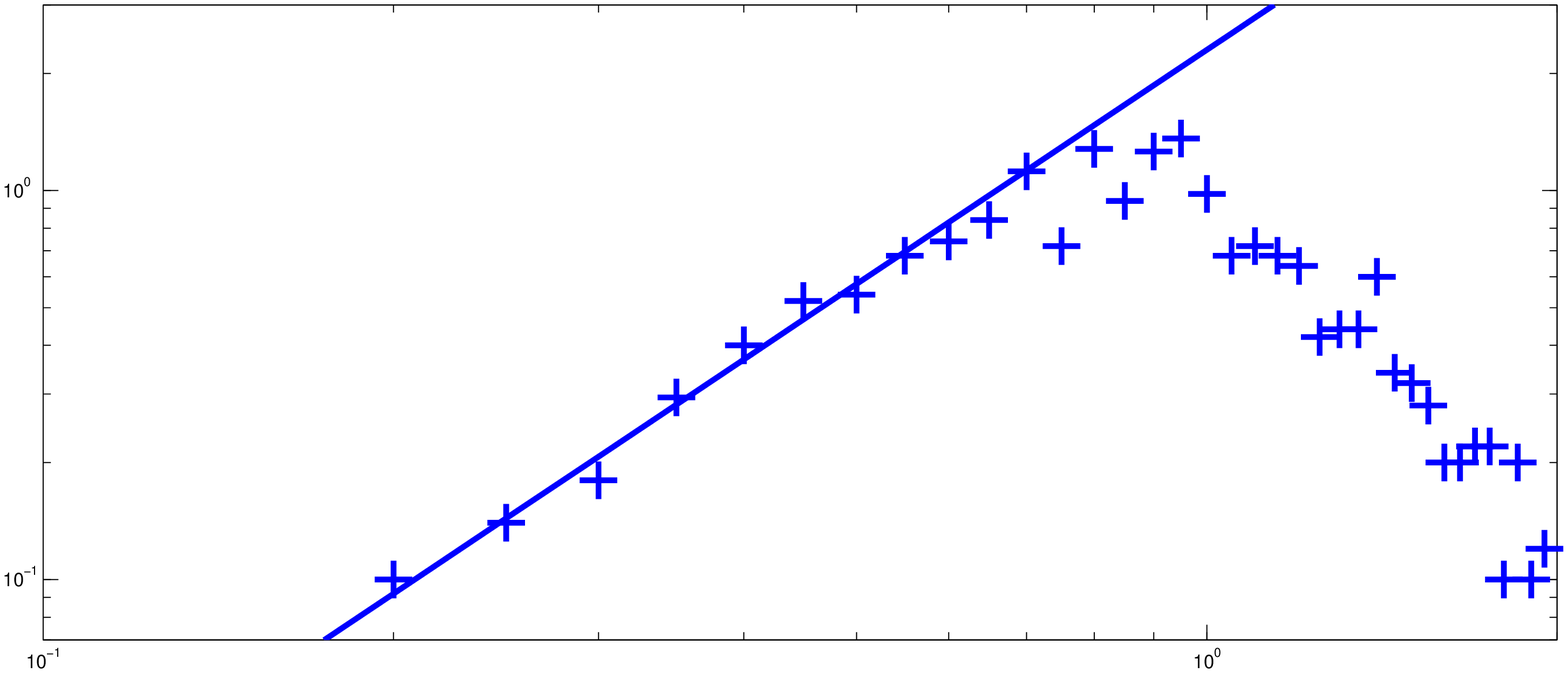}\\
\end{center}
  \caption{
  The probability density $p(D)$ evaluated for the measured data
  (crosses)
  is in a log-log scale compared with the function $2.3*D^2$ (full line). The agreement for small $D$
  is evident.
  }
  \label{tau3}
\end{figure}

We photographed the car queues in a front of the red light. The
photographes were taken all from one spot and at the same daytime.
The clearances were finally obtained by digitalization. Altogether
we extracted 1000 car distances from one particular crossing in the
city of Hradec Kralove (Czech republic) and evaluated the
corresponding probability density $p(D)$. Similar measurement has
been done also on several crossings in Prague - see \cite{krbalek}.
If there is a linear correlation between the stopping distance and
the estimated time to contact $\tau$ the obtained distance density
$p(D)$ should behave as $\approx D^2$ for small $D$. The result for
small distances is plotted on the Figure \ref{tau3}. It shows a very
nice agreement with that assumption.

To show that the $\tau$ mechanism is generic leading to a distance
distribution that is independent on the objects (the object can be a
man, animal or a car) we divide the further observations into two
categories: the first contains sedentary objects and the second
object moving in a dense environment.

Let us start with the the clearance distribution obtained for cars
parked in parallel and for birds perching on a power line. In both
cases the "parking segment" is full, i.e. there is not a free space
to place an additional participant. We have argued that under this
conditions the resulting distribution is invariant under the
transform (\ref{cnewdistances3}) with the parameter $g$ in
(\ref{distrib2}) fixed to $3$. The "parking" segments under
consideration were long and containing a large number of objects. So
$N>>1$, and the constrain (\ref{total}) does not play a substantial
role. In this case  $p(D)$ equals to $\Gamma(3,1,D)$.

To verify the prediction we measured the bumper to bumper distances
between cars parked in the center of Hradec Kralove (Czech
Republic). The street was located in a place with large parking
demand and usually without any free parking lots. Moreover it  was
free of any dividing elements, side ways and so on. Altogether we
measured 700 spacings under this conditions.

\begin{figure}
\begin{center}
  \includegraphics[height=9cm,width=15cm]{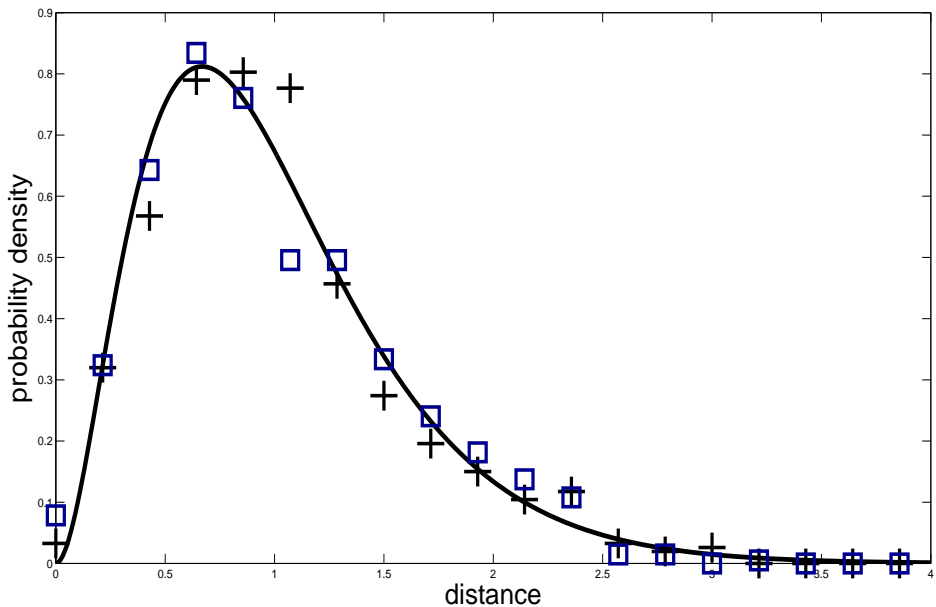}\\
\end{center}
  \caption{The probability density of the distances between the parked cars
  (crosses)
  and perching starlings (squares) is compared with the prediction of the
  theory (full line). The mean distance is normalized to 1.
  }
  \label{birds}
\end{figure}

For the birds we photographed flocks of starlings resting on the
power line during their flight to the south. The line was "full",
i.e. other starlings from the flock were forced to use another line
to perch. The bird-to-bird distances were obtained by a simple
digitalization - altogether 1000 bird spacings. After scaling the
mean distance to 1 the results were plotted on the Figure
\ref{birds} and compared with the prediction (\ref{distrib2}).

The probability distributions resulting from these data seems to be
(up to the statistical fluctuations) identical  and  in a good
agreement with the model prediction. This is amazing since the used
"hardware" is fully different. The underlying psychophysical
mechanism for the time to contact  estimation, is, however,
identical. (For the experimental results concerning the relevance of
$\tau$ for the space perception of pigeons see \cite{pigeon}.)

Let us now pass to the traffic streams, i.e. to a situation when the
objects are collectively moving. We assume a dense traffic. The
distances between the moving objects are small and have to be
constantly controlled to avoid  possible collisions.  We use the
mapping (\ref{cnewdistances4}).

Outside the safety margin the collision avoidance is assumed to be
based on  $\tau$. This means that the distribution of the distances
$D_k-d$ follows the mapping (\ref{cnewdistances3}) with the same
probability density $q(a)$ as in the "parking" case. So the
distributions of $D_k-d$ should be universal.

In order to verify this hypothesis we organized a simple experiment
with pedestrians in a narrow corridor. Using two light gates we
measured their velocity and the time interval that elapsed between
two subsequent walkers. This enables us to reconstruct the mutual
distances and evaluate the distance probability density. The same
device and method was used also for a sheep herd moving through an
aisle between two near yards. The third source of data are cars
moving on a highway in a dense traffic. The velocity and time stamps
of the individual cars were obtained by induction loops placed below
the roadway.

\begin{figure}
\begin{center}
  \includegraphics[height=9cm,width=15cm]{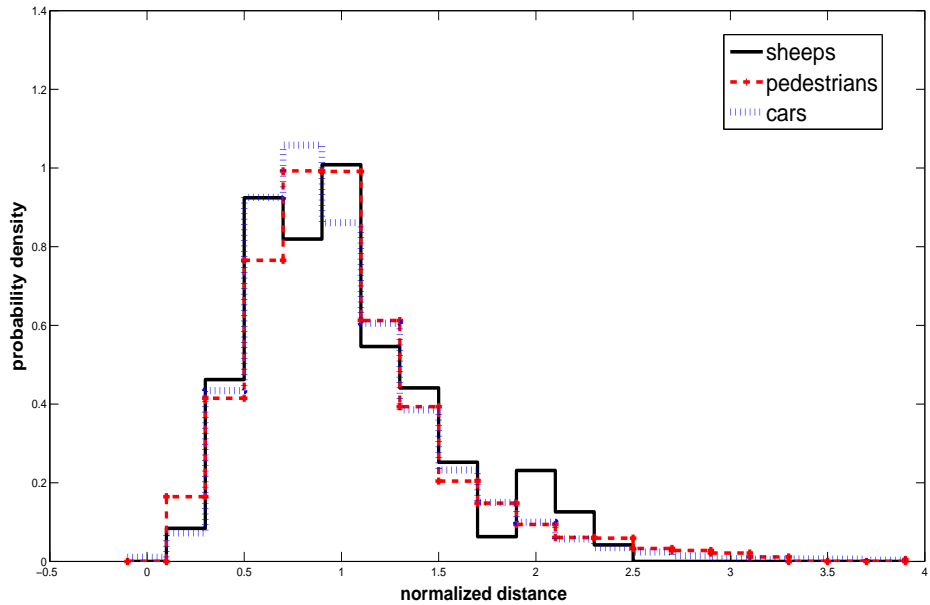}\\
\end{center}
  \caption{The probability density of the distances between the cars on
  highway in a high density regime, pedestrians walking in a narrow corridor and
  sheeps moving in an aisle between two barriers is plotted. The minimal distance on each particular data
  set is subtracted and the mean distance is normalized to 1.
  }
  \label{streams}
\end{figure}

The safety margins $d$ are clearly different in these three cases
and we removed them by subtracting the minimal distance from the
given data set. The mean distance was finally normalized to 1. The
result is plotted on the figure \ref{streams}

Again: the resulting distance distribution is universal and in
agreement with the theory based on $\tau$ hypothesis.

To summarize we have demonstrated that the clearance between objects
(cars,pedestrians,birds and sheeps) is largely universal. This
surprising observation can be understood as a consequence of an
universal distance controlling  mechanism shared by human and
animals.

{\bf Acknowledgement:} The research was supported by the Czech
Ministry of Education within the project  LC06002.  The help of the
PhD. students of the Department of Physics, University Hradec
Kralove who collected the majority of the data is gratefully
acknowledged. The help of Shinya Okazaki which was responsible for
the traffic light data is also gratefully acknowledged.

\end{document}